\documentclass[10pt,final,conference,letterpaper,romanappendices,twocolumn]{IEEEtran}
\usepackage{cite}
\usepackage{epsfig}
\usepackage{enumerate}
\usepackage{amsthm}
\usepackage{amsmath,amssymb,amsfonts,amstext,amsbsy,amsopn,dsfont}
\usepackage{cases}
\usepackage{bigints}
\usepackage{sublabel}
\usepackage{array}

\newtheorem{theorem}{\textbf{Theorem}}

\newtheorem{lemma}{\textbf{Lemma}}

\newcommand{\defn}{\triangleq}
\newcommand{\dif}{\textmd{d}}

\begin{document}

\title{Coverage Analysis for Dense Heterogeneous Networks with Cooperative NOMA}

\author{Chun-Hung Liu, Di-Chun Liang, Po-Chia Chen and Jie-Ru Yang\\Department of Electrical and Computer Engineering \\National Chiao Tung University, Hsinchu, Taiwan\\ e-mail: chungliu@nctu.edu.tw}

\maketitle

\begin{abstract}
In a heterogeneous cellular network (HetNet) consisting of $M$ tiers of densely-deployed base stations (BSs), consider that each of the BSs in the HetNet that are associated with multiple users is able to simultaneously schedule and serve two users in a downlink time slot by performing the (power-domain) non-orthogonal multiple access (NOMA) scheme. This paper aims at the preliminary study on the downlink coverage performance of the HetNet with the \textit{non-cooperative} and the proposed \textit{cooperative} NOMA schemes. First, we study the coverage probability of the NOMA users for the non-cooperative NOMA scheme in which no BSs are coordinated to jointly transmit the NOMA signals for a particular cell and the coverage probabilities of the two NOMA users of the BSs in each tier are derived. We show that the coverage probabilities can be largely reduced if allocated transmit powers for the NOMA users are not satisfied with some constraints. Next, we study and derive the coverage probabilities for the proposed cooperative NOMA scheme in which the void BSs that are not tagged by any users are coordinated to enhance the far NOMA user in a particular cell. Our analyses show that cooperative NOMA can significantly improve the coverage of all NOMA users as long as the transmit powers for the NOMA users are properly allocated.
\end{abstract}

\section{Introduction}
In traditional cellular networks, orthogonal multiple access (OMA) schemes, such as frequency
division multiple access (FDMA), time division multiple access (TDMA) and code division multiple access (CDMA), are able to successfully suppress a large amount of co-channel interference so that the signal-to-interference power ratio  (SIR) at the receiver side can be enhanced remarkably. However, enhancing SIR via OMA is not the most efficient/effective method to improve the spectrum efficiency of a wireless link dominated by network co-channel interference according to the fundamental result of the multiuser capacity region \cite{DTPV05}\cite{LDBWYYSHCLIZW15}. To fulfill the huge throughput growth anticipated in 5G cellular networks under the pressing pressure of spectrum crunch, non-orthogonal multiple access (NOMA) that is able to reduce the complexity in resource allocation and user scheduling as well as make the scarce spectrum resource be utilized and shared in a more efficient means has gained a lot of attentions recently \cite{ABYSYKALAHTN13,LDBWYYSHCLIZW15,SMRNAOADKSK17}. 

It is well known that in a downlink cellular network the power-multiplexing NOMA scheme\footnote{The NOMA scheme in this paper is a multiplexing scheme in the power domain \cite{LDBWYYSHCLIZW15,SMRNAOADKSK17}, i.e., different downlink users are allocated different powers based on their channel gain conditions, whereas other code-domain-multiplexing NOMA schemes are beyond the scope of this work.} that adopts successive interference cancellation  (SIC) to perfectly cancel the multiuser interference always achieves a larger sum throughput (achievable sum rate) than the OMA scheme. However,  the coverage performance of the scheduled NOMA users definitely is weakened due to power reduction for each user. In an interference-limited cellular network, the coverage probability, i.e., the probability that the SIR of users in the network is higher than some predesignated threshold, actually dominates the sum downlink throughput of the scheduled NOMA users so that the downlink throughput cannot not be improved provided users' coverage performance is severely degraded. The coverage-degraded problem of the NOMA users becomes even much serious in a dense heterogeneous cellular network (HetNet) where a large amount of interference is generated by many different kinds of base stations densely deployed in the HetNet. 

The previous works on the study of the coverage/outage problem in a HetNet with the NOMA scheme are still very minimal and just few current works have studied the transmission performance of the NOMA scheme based on a simple single-cell model. For example, reference \cite{ZDZYPFHVP14} studied the performance of the outage and ergodic rate of the NOMA scheme in a single cell and showed that NOMA can achieve a higher sum rate whereas its rate gain in the low SNR region is not significant. In \cite{PXYYZDXDRS16}, a cooperative NOMA scheme was proposed to simultaneously transfer wireless information and power for users in a single cell and the outage probabilities for different user selection schemes were analyzed. These two works were not developed in a multi-cell network and thus how the outage and rate performance studied in them is impacted by multi-cell interference is not clear at all. 

To investigate the coverage performance of the NOMA users in a HetNet, in this paper we consider a multi-tier HetNet in which BSs that are associated by multiple users can perform the NOMA scheme to serve their two scheduled users. First we study the coverage performance of the \textit{non-cooperative} NOMA scheme in which no BSs are coordinated to enhance the signals of the far NOMA user in a particular cell\footnote{This cooperative NOMA scheme is motivated by our recent work in the load-aware coordinated multi-point joint transmission (JT-CoMP) in \cite{CHLPCC17}.}.  The coverage probabilities of the NOMA users are derived and they are shown to be seriously degraded if the transmit power is not properly allocated to the NOMA users. In order to improve the coverage performance of the NOMA users, we propose the cooperative NOMA scheme in which all the void BSs that are not associated with any users are coordinated to jointly transmit the signals of the far NOMA user in a particular cell. Our main finding indicates that cooperative NOMA is able to significantly improve all NOMA users as long as the transmit power allocated to the far NOMA user is higher than some threshold. Also, we show that there exists an optimal power allocation that maximizes the average of the coverage probabilities of the users for the non-cooperative and cooperative NOMA schemes.

\section{System Model and Assumptions}
Consider a large-scale interference-limited heterogeneous cellular network (HetNet) on the plane $\mathbb{R}^2$ in which there are $M$ different types of base stations (BSs) (e.g., macrocell, microcell, picocell BSs, etc.) and the BSs of each type are referred as a tier of the HetNet. Specifically, we assume that the tier-$m$ BSs form an independent homogeneous Poisson point process (PPP) of intensity $\lambda_m$ and they are denoted by set $\Phi_m$ that can be explicitly written as
\begin{align}
\Phi_m\defn \{X_{m,i}\in\mathbb{R}^2 : i\in\mathbb{N}_+\}, \,\,m\in\mathcal{M},
\end{align}
where set $\mathcal{M}\defn\{1,2,\ldots,M\}$ and $X_{m,i}$ denotes BS $i$ in the $m$th tier and its location. Also, all users are assumed to form an independent PPP of intensity $\mu$ and they are denoted by set $\mathcal{U}$ given as
\begin{align}
\mathcal{U}\defn \{U_j\in\mathbb{R}^2 : j\in\mathbb{N}_+\} ,
\end{align} 
where $U_j$ represents user $j$ and its location. For simplicity and tractability in analysis, in this paper all users are assumed to associate with their nearest BS\footnote{To make the following analysis much more tractable, in this work no user association bias is used for each tier. More general user association schemes with a random bias for each tier, such as maximum received-power association and energy-efficient user association, can be referred to references \cite{CHLLCW16,CHLKLF16}.}. Let $\mathcal{C}_{m,i}$ be the voronoi-tessellated cell of BS $X_{m,i}$ and $\mathcal{U}_{m,i}\defn\{U_j\in\mathcal{C}_{m,i}: j\in\mathbb{N}_+\}$ is the set of the users associating with BS $X_{m,i}$.

 Let $N$ be the number of the users associating with a BS and its probability mass function (pmf) can be accurately found based on Lemma 1 in our previous work \cite{CHLKLF16} as
\begin{align}\label{Eqn:pmfNumUsers}
\mathbb{P}[N=n]\approx\frac{\Gamma(n+\frac{7}{2})}{n! \Gamma(\frac{7}{2})}\left(\frac{2}{7}L\right)^{n}\left(\frac{1}{1+\frac{2}{7}L}\right)^{n+\frac{7}{2}},
\end{align}   
where $L\defn \frac{\mu}{\sum_{m=1}^{M}\lambda_m}$ is called the cell load of a BS and it represents the mean number of users associating with a BS. Thus, \eqref{Eqn:pmfNumUsers} indicates  $\mathbb{P}[N=0]=(1+\frac{2}{7}L)^{-\frac{7}{2}}$, which is called void cell (BS) probability and denotes the probability that a BS is not associated by any users (i.e., it is void). In other words, the non-void probability of a BS can be readily written as
\begin{align}
q\defn 1-\mathbb{P}[N=0]\approx1-\left(1+\frac{2}{7}L\right)^{-\frac{7}{2}}.
\end{align}
Note that the non-void probability $q$ (void probability $\mathbb{P}[N=0]$) is small (large) when the user intensity is not much smaller than the total intensity of BSs, for example, a dense HetNet with a high intensity (density) of deployed BSs. As a result, the intensity of the void BSs, $(1-q)\sum_{m=1}^{M}\lambda_m$ would not be small, which means the void cell phenomenon that is usually overlooked in the literature needs to be carefully considered in the interference model of a dense HetNet since those void BSs actually do not generate interference \cite{CHLLCW16}. 

In this paper, our focus is to study the downlink coverage performance of a BS when all BSs with at least two tagged users can perform the non-orthogonal multiple access (NOMA) scheme to simultaneously transmit different data to different users over the same frequency band. The NOMA scheme considered in this paper is performed in the power domain, i.e., users are allocated different transmit powers while transmitting their data according to their channel conditions \cite{LDBWYYSHCLIZW15}, and users are able to performance successive interference cancellation (SIC) so as to decode its own data. To tractably and simply analyze the downlink coverage of the users scheduled by a tier-$m$ BS, we consider a downlink NOMA scheme in which the BSs with two or more users arbitrarily schedule any two users for doing NOMA transmission, whereas the BSs associated by only one user, of course, use their full power to deliver the data to their sole user. Furthermore, all BSs and users are assumed to be equipped with a single antenna. 

\section{Downlink Coverage Analysis with Non-cooperative NOMA}\label{Sec:CovProbNonCoopNOMA}
In this section, we would like to study the downlink coverage performance of a tier-$m$ BS performing the non-cooperative NOMA scheme if the tier-$m$ BS has multiple users. Since each BS with multiple users only schedule two users for NOMA transmission in each downlink time slot, consider $U_{j}, U_{k}\in\mathcal{U}_{m,i}$ for $j\neq k$ are two users scheduled by BS $X_{m,i}$. Without loss of generality, suppose BS $X_{m,i}$ is located at the origin\footnote{According to the Slivnyak theorem \cite{DSWKJM13}, the statistical properties evaluated at any particular point in homogeneous PPPs are the same.} and $\|U_j\|<\|U_k\|$ where $\|X-Y\|$ is the Euclidean distance between nodes $X$ and $Y$, i.e., $U_j$ is closer to BS $X_{m,i}$ than $U_k$. Hence, $X_{m,i}$ allocates more power to transmit the data of $U_k$ in order to make $U_j$ successfully perform SIC with a higher probability. The \textit{desired} signal-to-interference power ratio (SIR) for $U_j$ or $U_k$  is defined as
\begin{align}\label{Eqn:DSIR}
\gamma_{m,l}\defn \frac{P_{m,l}H_{m,l}}{\|U_{l}\|^{\alpha}I_{m,l}},
\end{align}
where $l\in\{j,k\}$, $\alpha>2$ is the pathloss exponent, $P_{m,l}= P_m[(1-\beta_m)\mathds{1}(l=j)+\beta_m\mathds{1}(l=k)]$ is the transmit power for user $U_l$, $\beta_m\in[0,1]$ is the power allocation factor of a tier-$m$ BS with two NOMA users, $\mathds{1}(\mathcal{E})$ stands for the indicator function that is equal to one if event $\mathcal{E}$ is true and zero otherwise, $I_{m,l}$ is the interference received by $U_l$ and it is given by
\begin{align*}
I_{m,l}\defn \sum_{k,j\in\Phi\setminus X_{k,i}} V_{k,j}P_kH_{k,j} \|X_{k,j}-U_l\|^{-\alpha}
\end{align*}  
in which $\Phi\defn\bigcup_{m=1}^M \Phi_m$, $V_{m,j}\in\{0,1\}$ is a Bernoulli random variable that is one if BS $X_{m,j}$ is non-void and zero otherwise, and $H_{m,j}$ is the Rayleigh fading channel gain from BS $X_{m,j}$ to user $U_l$. Throughput this paper, all channel gains are assumed to be i.i.d. exponential random variables with unit mean and variance, i.e., $H_{m,j}\sim\mathrm{Exp}(1)$ for all $m\in\mathcal{M}$ and $j\in\mathbb{N}_+$, and the shadowing effect on all channels is ignored in order to facilitate the following analysis.  

According to \eqref{Eqn:DSIR}, we define the coverage probability of the near user $U_j$ as
\begin{align}
\overleftarrow{\rho}_{m}(\beta_m,\theta)&\defn \mathbb{P}\left[\frac{P_{m,k}H_{m,j}}{P_{m,j}H_{m,j}+\|U_j\|^{\alpha}I_{m,j}}\geq\theta,\gamma_{m,j}\geq \theta\right] \nonumber\\
&=\mathbb{P}\left[\frac{\left(\frac{\beta_m}{1-\beta_m}\right)\gamma_{m,j}}{\gamma_{m,j}+1}\geq \theta,\gamma_{m,j}\geq \theta\right]\label{Eqn:CovProbUj}
\end{align}
in which $\theta>0$ is the SIR threshold for successful decoding, and define the coverage probability of the far user $U_k$ as
\begin{align}
\overrightarrow{\rho}_{m}(\beta_m,\theta)&\defn\mathbb{P}\left[\frac{H_{m,k}\beta_mP_m\|U_k\|^{-\alpha}}{(1-\beta_m)H_{m,k}P_m\|U_k\|^{-\alpha}+I_{m,k}}\geq\theta\right]\nonumber\\
&=\mathbb{P}\left[\frac{\gamma_{m,k}}{(\frac{1-\beta_m}{\beta_m})\gamma_{m,k}+1}\geq\theta\right]. \label{Eqn:CovProbUk}
\end{align}
Note that $\overleftarrow{\rho}_m(\beta_m,\theta)$ has to include the event of successfully decoding the signal of $U_k$ since $U_j$ has to decode $U_k$'s signal first and then perform SIC to decode its own signals.  The explicit results of $\overleftarrow{\rho}_{m}(\beta_m,\theta)$ and $\overrightarrow{\rho}_{m}(\beta_m,\theta)$ are shown in the following theorem.
\begin{theorem}\label{Thm:CovProbNOMA}
If power allocation factor $\beta_m\in\left(\frac{\theta}{1+\theta},1\right]$, then the accurate coverage probabilities of the two scheduled users associating with a tier-$m$ BS, as defined in \eqref{Eqn:CovProbUj} and \eqref{Eqn:CovProbUk}, can be expressed as
\begin{align}\label{Eqn:CovProbUserj}
\overleftarrow{\rho}_{m}(\beta_m,\theta)&\approx \frac{2}{2+q\ell_m(\hat{\theta}_m)},\\
\overrightarrow{\rho}_{m}(\beta_m,\theta)&\approx \frac{2}{\left[1+q\ell_m(\tilde{\theta}_m)\right]\left[2+q\ell_m(\tilde{\theta}_m)\right]},\label{Eqn:CovProbUserk}
\end{align}
where $\hat{\theta}_m\defn \max\{\frac{\theta}{\beta_m(1+\theta)-\theta},\frac{\theta}{1-\beta_m}\}$, $\tilde{\theta}_m\defn \frac{\theta}{\beta_m(1+\theta)-\theta}$ and  $\ell_m(x)$ is given by
\begin{align*}
\ell_m(x)\defn\sum_{k=1}^{M} \vartheta_k \left(\frac{xP_k}{P_m}\right)^{\frac{2}{\alpha}}\left(\frac{2\pi/\alpha}{\sin(2\pi /\alpha)}-\int_{0}^{(\frac{P_m}{xP_k})^{\frac{2}{\alpha}}}\frac{\dif t}{1+t^{\frac{\alpha}{2}}}\right)
\end{align*}
and $\vartheta_k\defn\frac{\lambda_k}{\sum_{m=1}^{M}\lambda_m}$.
\end{theorem}
\begin{IEEEproof}
See Appendix \ref{App:ProfCovProbNOMA}.
\end{IEEEproof}
The results in Theorem \ref{Thm:CovProbNOMA} reveal a few important implications regarding how to allocate the powers for the two NOMA users. First, as shown in the proof of Theorem \ref{Thm:CovProbNOMA}, the two NOMA users cannot decode their own signals almost surely if $\beta_m\in[0,\frac{\theta}{\theta+1}]$ because the signals of the far user cannot be decoded even when there is no interference (i.e., decoding the desired signals of each user directly fails due to the self-interference between the NOMA users.). Thus, $\beta_m$ needs to be properly chosen in the interval $(\frac{\theta}{\theta+1},1]$ in order to make the NOMA scheme work well. Second, if $\beta_m\in(\frac{1+\theta}{2+\theta},1]$, the coverage probability of the near user is always superior to that of the far user since in this case the performance of decoding the signals of the far user dominates the coverage performance of the two users. Third, the optimal value of $\beta_m$ that maximizes the average of the two coverage probabilities can be found by formulating the following optimization problem for a given $\theta>0$:
\begin{align}\label{Eqn:OptCovProb}
\begin{cases}
\text{maximize }_{\beta_m}\frac{1}{2}\left[\overleftarrow{\rho}_{m}(\beta_m,\theta)+\overrightarrow{\rho}_{m}(\beta_m,\theta)\right]\\
\text{subject to }\frac{\theta}{1+\theta}<\beta_m\leq 1
\end{cases}.
\end{align}
As shown in the following lemma, there exists an optimal value of $\beta_m$ that maximizes the average of the two coverage probabilities in \eqref{Eqn:OptCovProb}.
\begin{lemma}\label{Lem:OptimalPowerAllocNonCooPNOMA}
For a given $\theta>0$, there exists a $\beta^{\star}_m\in(\frac{\theta}{\theta+1},1]$ that maximizes $\frac{1}{2}\left[\overleftarrow{\rho}_m(\beta_m,\theta)+\overrightarrow{\rho}_m[\beta_m,\theta)\right]$. 
\end{lemma}
\begin{IEEEproof}
Since $\overleftarrow{\rho}_m(\beta_m,\theta)+\overrightarrow{\rho}_m(\beta_m,\theta)$ is continuous over $[\frac{\theta}{\theta+1},1]$ and we also know the following
\begin{align*}
&\lim_{\beta_m\rightarrow 1}\overleftarrow{\rho}_m(\beta_m,\theta)+\overrightarrow{\rho}_m(\beta_m,\theta)<1,\\
&\lim_{\beta_m\rightarrow \frac{\theta}{\theta+1}}\overleftarrow{\rho}_m(\beta_m,\theta)+\overrightarrow{\rho}_m(\beta_m,\theta)=0.
\end{align*}
Hence $\overleftarrow{\rho}_m(\beta_m,\theta)+\overrightarrow{\rho}_m(\beta_m,\theta)$ is bounded over $[\frac{\theta}{\theta+1},1]$. Thus, for any $\beta_m\in(\frac{\theta}{\theta+1},1]$ we know $\overleftarrow{\rho}_m(\beta_m,\theta)+\overrightarrow{\rho}_m(\beta_m,\theta)\in(0,1)$ so that there must exist an optimal value of $\beta^{\star}_m$ that maximizes $\overleftarrow{\rho}_m(\beta_m,\theta)+\overrightarrow{\rho}_m(\beta_m,\theta)$ for a given $\theta>0$ based on the Weierstrass theorem \cite{DPB16}. 
\end{IEEEproof}  
Although the optimal value $\beta^{\star}_m$ may not be found in closed-form due to the complexity of the function $\ell_m(\cdot)$, it can be easily found by numerical techniques.  

\section{Downlink Coverage Analysis with Cooperative NOMA}\label{Sec:CovProbCoopNOMA}
In the previous section, the coverage probabilities for the non-cooperative NOMA scheme are investigated and they are shown to be severely impacted whether or not power allocation between NOMA users and SIC are appropriately and successfully performed. To reinforce the performance of power allocation and SIC, we propose a cooperative NOMA scheme in which all void BSs can be perfectly coordinated to jointly transmit the signals of the far user among the two scheduled NOMA users in a particular cell. In the following, we will also investigate the coverage probability performance of the users for this proposed cooperative NOMA scheme. 
 
By following the same analysis approach in the previous section, consider BS $X_{m,i}$ located at the origin can perform the cooperative NOMA scheme whenever it is able to schedule two users $U_j$ and $U_k$ for transmission. Since all the void BSs are coordinated to transmit the signals of the far user $U_k$, the coverage probability of the near user $U_j$ defined in \eqref{Eqn:CovProbUj} for the cooperative NOMA scheme can be explicitly expressed as
 \begin{align*}
 &\hspace{-0.1in}\overleftarrow{\rho}_m(\beta_m,\theta)=\mathbb{P}\left[\frac{\left(\frac{\beta_m}{1-\beta_m}\right)\gamma_{m,j}+\frac{S^c_m}{I_{m,j}}}{\gamma_{m,j}+1}\geq\theta, \gamma_{m,j}\geq \theta\right]
 \end{align*}
 \begin{align}
 &\hspace{-0.1in}=\mathbb{P}\left[\gamma_{m,j}\geq\max\left\{\frac{(1-\beta_m)}{\beta_m(1+\theta)-\theta}\left(\theta-\frac{S^c_m}{ I_{m,j}}\right),\theta\right\}\right],\label{Eqn:CovProbCoopNOMAUj}
 \end{align} 
where $S^c_m\defn  \sum_{l,n\in\Phi\setminus X_{m,i}} (1-V_{l,n})P_lH_{l,n} \|X_{l,n}-U_j\|^{-\alpha}$ denotes the sum of the signal powers of the far user $U_k\in\mathcal{C}_{m,i}$ from all coordinated void BSs that are used to coherently transmit the signals of the far user $U_k\in\mathcal{C}_m$ and $\beta_m\in(\frac{\theta}{\theta+1},1]$. The coverage probability of the far user $U_k$ can be written as
\begin{align}\label{Eqn:CovProbCoopNOMAUk}
\overrightarrow{\rho}_m(\beta_m,\theta) \defn& \mathbb{P}\left[\frac{\beta_m P_m\|U_k\|^{-\alpha}+S^c_m}{(1-\beta_m)P_m\|U_k\|^{-\alpha}+I_{m,k}}\geq\theta\right]\nonumber\\
=&\mathbb{P}\left[\gamma_{m,k}\geq\beta_m\tilde{\theta}_m\left(1-\frac{S^c_m}{\theta I_{m,k}}\right)\right]
\end{align} 
for $\beta_m\in(\frac{\theta}{1+\theta},1]$. The analytical results of $\overleftarrow{\rho}_m(\beta_m,\theta) $ in \eqref{Eqn:CovProbCoopNOMAUj} and $\overrightarrow{\rho}_m(\beta_m,\theta)$ in \eqref{Eqn:CovProbCoopNOMAUk} can be derived and shown in the following theorem.

\begin{theorem}\label{Thm:CovProbCooPNOMA}
If the cooperative NOMA scheme is performed in the HetNet, the coverage probability in \eqref{Eqn:CovProbCoopNOMAUj} for $\beta_m\in[\frac{1+\theta}{2+\theta},1]$ is accurately given by
\begin{align}\label{Eqn:DerCoverProbCoopNOMAUserj}
\overleftarrow{\rho}_{m}(\beta_m,\theta)\approx &\frac{2}{2+q\ell_m\left(\frac{\theta}{1-\beta_m}\right)},
\end{align}
and the coverage probability in \eqref{Eqn:CovProbCoopNOMAUk} is accurately found as
\begin{align}\label{Eqn:DerCoverProbCoopNOMAUserk}
\overrightarrow{\rho}_m(\beta_m,\theta)\approx \frac{2}{\left[1+Q_m\left(\tilde{\theta}_m,\frac{\tilde{\theta}_m}{\theta}\right)\right]\left[2+Q_m\left(\tilde{\theta}_m,\frac{\tilde{\theta}_m}{\theta}\right)\right]},
\end{align}
where  $Q_m(x,y)$ for $x,y\in\mathbb{R}_{++}$ is defined as
\begin{align*}
Q_m(x,y)=[q\ell_m(x)+(1-q)\widetilde{\ell}_m(y)]^+
\end{align*}
in which $(x)^+\defn\max\{x,0\}$ and $\widetilde{\ell}_m(y)$ is defined as
\begin{align}
\widetilde{\ell}_m(y) \defn \sum_{k=1}^{M} \vartheta_k\left(\frac{yP_k}{P_m}\right)^{\frac{2}{\alpha}}\int^{\infty}_{\left(\frac{P_m}{yP_k}\right)^{\frac{2}{\alpha}}}\mathbb{E}\left[1-e^{t^{-\frac{\alpha}{2}}H}\right]\dif t
\end{align}
\end{theorem}
\begin{IEEEproof}
See Appendix \ref{App:ProfCovProbCooPNOMA}.
\end{IEEEproof}

The coverage results in Theorem \ref{Thm:CovProbCooPNOMA} clearly indicate how the cooperative NOMA scheme improves the coverage probabilities of the two NOMA users. For the near user, if we compare \eqref{Eqn:DerCoverProbCoopNOMAUserj} with \eqref{Eqn:CovProbUserj}, the coverage probability in \eqref{Eqn:DerCoverProbCoopNOMAUserj} is higher than that in \eqref{Eqn:CovProbUserj} and it is dominated by the (desired) SIR of the near user that is not affected by the interference of the far user. This is because cooperative NOMA makes the first decoding process of the near user succeed almost surely. Also, the coverage probability of the far user is also enhanced if comparing \eqref{Eqn:DerCoverProbCoopNOMAUserk} with \eqref{Eqn:CovProbUserk} since $Q_m(\tilde{\theta},\tilde{\theta}/\theta)$ is definitely smaller than $q\ell_m(\tilde{\theta})$. Note that the coverage probabilities achieved by cooperative NOMA both increase as the user intensity reduces since more void BSs can be coordinated to improve the desired signal strength of the far user. On the contrary, the coverage performance of cooperative NOMA degrades to that of the non-cooperative NOMA as more and more users join the HetNet. Thus, here arises an interesting and important problem about how to maintain an appropriate cell load of the BSs in each tier (see \eqref{Eqn:pmfNumUsers}) so that there exists a good number of the void BSs that can be coordinated to perform the proposed cooperative NOMA scheme for a given user intensity. 

Note that the minimum required range for the power allocation factor $\beta_m$ in Theorem \ref{Thm:CovProbCooPNOMA}, i.e.,  $\beta\in[\frac{1+\theta}{2+\theta},1]$, is (slightly) higher than that posed in Theorem \ref{Thm:CovProbNOMA}. This required range of $\beta_m$ is obtained by facilitating the derivations of the coverage probabilities in Theorem \ref{Thm:CovProbCooPNOMA} when using the cooperative NOMA scheme. In fact, the cooperative probabilities essentially can be improved by cooperative NOMA for any $\beta_m\in(0,1)$. Most importantly, this range $\beta_m\in[\frac{1+\theta}{2+\theta},1]$ makes us able to realize that cooperative NOMA can achieve the coverage probability limit of each user with a high probability. Furthermore, there also exists an optimal value of $\beta_m$ such that the average of the two coverage probabilities in Theorem \ref{Thm:CovProbCooPNOMA} is maximized, as shown in the following lemma.
\begin{lemma}\label{Lem:OptimalPowerAllocCooPNOMA}
For a given $\theta>0$, there exists a $\beta^{\star}_m\in(0,1)$ that maximizes $\frac{1}{2}[\overleftarrow{\rho}_m(\beta_m,\theta)+\overrightarrow{\rho}_m(\beta_m,\theta)]$ where $\overleftarrow{\rho}_m(\beta_m,\theta)$ and $\overrightarrow{\rho}_m(\beta_m,\theta)$ are given in \eqref{Eqn:CovProbCoopNOMAUj} and \eqref{Eqn:CovProbCoopNOMAUk}, respectively. 
\end{lemma}
\begin{IEEEproof}
The proof is omitted here since it is similar to the proof of Lemma \ref{Lem:OptimalPowerAllocNonCooPNOMA}.
\end{IEEEproof}
Similarly, $\beta^{\star}_m$ may only be found by numerical techniques due to the complexity of function $\ell_m(\cdot)$, and we can expect that the optimal power allocation factor for cooperative NOMA is smaller than that for non-cooperative NOMA since joint transmission enhances the signal power of the far user so that allocating less power to the far user may not degrade the decoding performance at the two users. In the following, we will present some numerical results to validate our analytical results and findings in above.

\begin{table}[!t]
	\centering
	\caption{Network Parameters for Simulation}\label{Tab:SimPara}
	\begin{tabular}{|c|c|c|}
		\hline Parameter $\setminus$ BS Type& Macrocell & Picocell\\ 
		\hline Power $P_m$ (W) & 20 & 2\\ 
		\hline Intensity $\lambda_m$ (BSs/m$^2$) & $1.0\times 10^{-6}$ & $[\mu, 0.1\mu]$    \\ 
		\hline User Intensity $\mu$ (users/m$^2$) &\multicolumn{2}{c|}{$5\times 10^{-4}$} \\ 
		\hline SIR Threshold $\theta$ &\multicolumn{2}{c|}{1} \\ 
		\hline Pathloss Exponent $\alpha$ & \multicolumn{2}{c|}{4} \\ 
		\hline Power Allocation factor $\beta_m$ & \multicolumn{2}{c|}{$\frac{3}{4}$} \\ 
		\hline 
	\end{tabular} 
\end{table}

\section{Numerical Results}
\begin{figure}[!t]
\centering
\includegraphics[width=3.65in,height=2.5in]{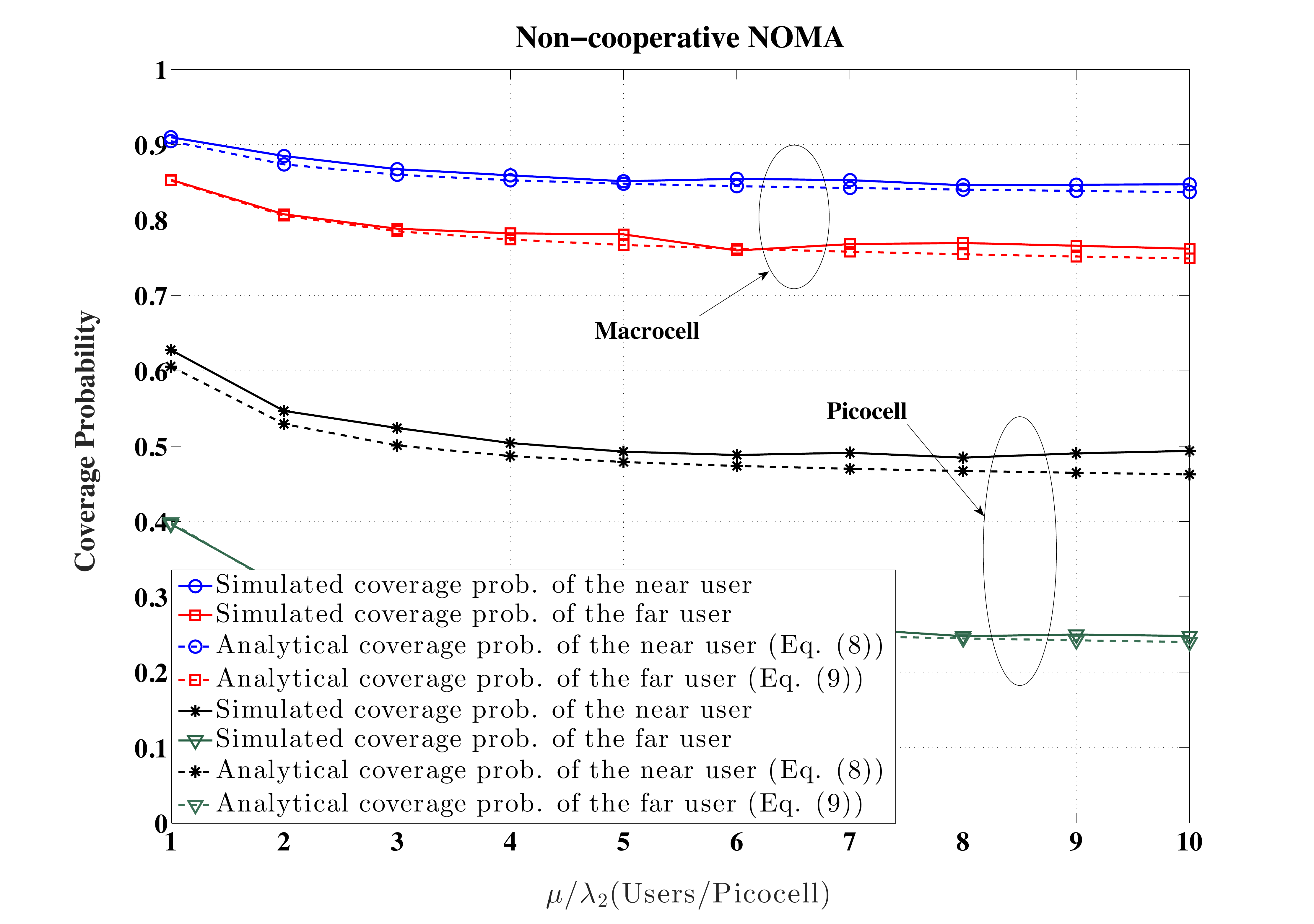}
\caption{Simulation results of the coverage probabilities for the non-cooperative NOMA scheme}
\label{Fig:CovprobNonCoopNOMA}
\end{figure}

In this section, some numerical results are provided to validate the coverage analysis with the non-cooperative and cooperative NOMA schemes in the previous section. We consider a two-tier HetNet consisting a tier of macrocell BSs and a tier of picocell BSs. The network parameters for simulation are listed in Table \ref{Tab:SimPara}. The simulation results of the coverage probabilities for the non-cooperative NOMA scheme are shown in Fig. \ref{Fig:CovprobNonCoopNOMA}. As can be observed, the analytical results are pretty close to their corresponding simulated results, which validates our previous analysis is fairly correct and accurate. Also, we can see that the coverage probabilities of the user in picocells are significantly smaller than those in the marcocells owing to the large transmit power of the marcocell BSs. The coverage probabilities essentially decrease as the user intensity increases since the interference increases due to the increase in the non-void probability and thus the average number of the non-void BSs in the network increases. Accordingly, all coverage probabilities eventually coverage to a constant value as the user intensity goes to infinity. Since the simulation setting leads to $\beta_m>\frac{1+\theta}{2+\theta}$, the near user almost surely has a better coverage than the far user based on our previous discussion in Section \ref{Sec:CovProbNonCoopNOMA} and the simulation results in Fig. \ref{Fig:CovprobNonCoopNOMA} support this point as well. Hence, power allocation surely dominates the SIR performance of the non-cooperative NOMA scheme.

\begin{figure}[!t]
	\centering
	\includegraphics[width=3.65in, height=2.5in]{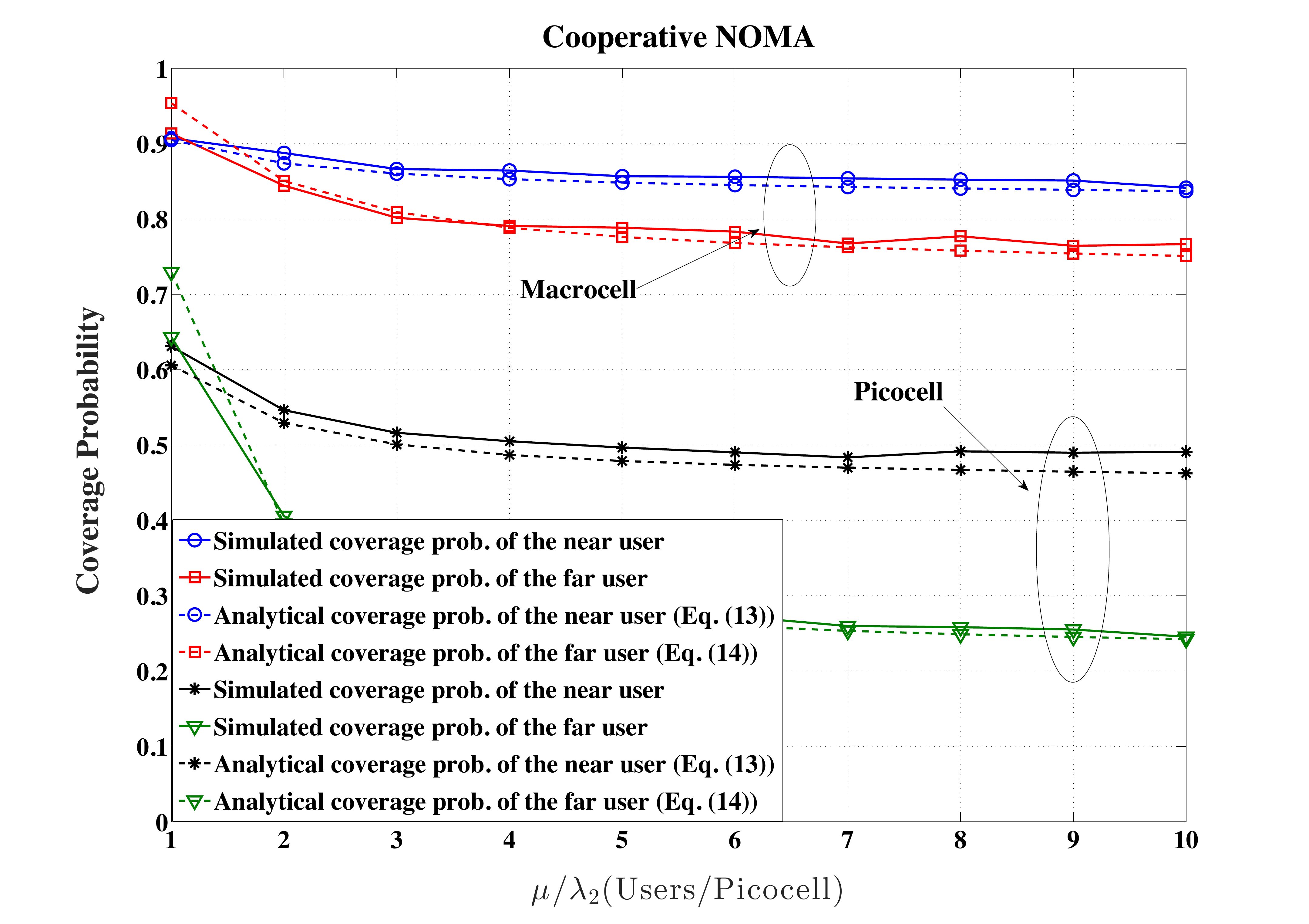}
	\caption{Simulation results of the coverage probabilities for the cooperative NOMA scheme}
	\label{Fig:CovprobCoopNOMA}
\end{figure}

The simulation results of the coverage probabilities for the cooperative NOMA scheme are shown in Fig. \ref{Fig:CovprobCoopNOMA}. By comparing Figs. \ref{Fig:CovprobNonCoopNOMA} and \ref{Fig:CovprobCoopNOMA}, we see that cooperative NOMA is able to significantly improve the coverage performance of the NOMA users when the user intensity is not much larger than the intensity of all the BSs, especially for the far users in the picocell BSs. However, the coverage probabilities reduce as the user intensity increases because there are not many void BSs that can be coordinated to perform the cooperative NOMA scheme. The coverage probability of the near user is not benefited much by cooperative NOMA since the near user already performs SIC almost perfectly in the power allocation with $\beta_m>\frac{1+\theta}{2+\theta}$ even without the help from joint transmission. Thus, the near user has a worse coverage than the far user only when the user intensity is not larger than the sum of the BS intensities.

\section{Concluding Remarks}
In the downlink transmission, a BS that performs the NOMA scheme to simultaneously serve two users can improve its downlink coverage if SIC is perfect at the use side. However, due to channel fading and inter-cell co-channel interference, SIC may fail at the near user end so that NOMA may not provide a good coverage performance to all users at the same time. In this paper, we first show that the non-cooperative NOMA scheme can severely degrade the coverage performance of the BSs in a HetNet if the powers for the NOMA users are not allocated properly. In order to significantly improve the SIC and NOMA, we propose a cooperative  NOMA scheme in which all void BSs can do joint transmission to enhance the signal power of the far NOMA user in a particular. The proposed cooperative NOMA scheme indeed significantly enhances the coverage of the BSs, especially in a dense network with a moderate user intensity. In our future work, we will study the coverage and throughput performance of the NOMA scheme in the case that more than two users can be served by NOMA in a HetNet.

\appendix
\subsection{Proof of Theorem \ref{Thm:CovProbNOMA}}\label{App:ProfCovProbNOMA}
To obtain the tractable result on the coverage probabilities, we neglect the location correlations between non-void BSs due to user association  and assume that the non-void BSs in the $m$th tier can be approximated by a thinning PPP of intensity $q\lambda_m$. Such an approximation has been shown to be still very accurate in modeling the locations of the non-void BSs\cite{CHLLCW15,CHLLCW16}. Hence, $\mathbb{P}[\gamma_{m,l}\geq \theta]$ can be accurately approximated by
\begin{align*}
\mathbb{P}[\gamma_{m,l}\geq \theta]&=\mathbb{P}\left[H_{m,l}\geq \theta\frac{ I_{m,l}\|U_l\|^{\alpha}}{P_{m,l}}\right]\\
&=\mathbb{E}\left[\exp\left(-\frac{\theta\|U_{l}\|^{\alpha}}{P_{m,l}}I_{m,l}\right)\right]\\
&\stackrel{(a)}{\approx}\mathbb{E}\left[\exp\left(-\pi \lambda_{\Sigma}\|U_l\|^2q\ell_m\left(\frac{\theta P_m}{P_{m,l}}\right)\right)\right],
%&\stackrel{(b)}{\approx}\int_{0}^{\infty}\pi\lambda_{\Sigma} e^{-\pi x\left[q\sum_{k=1}^{M}\lambda_k\ell\left(\frac{P_k\theta}{P_{m,l}},\frac{2}{\alpha}\right)+\lambda_{\Sigma}\right]}\dif x,
\end{align*}
where $(a)$ follows from the derivations of Proposition 1 in \cite{CHLLCW16} by using the fact that the associated non-void BSs still form a thinning independent PPP with intensity $q\lambda_{\Sigma}$ where $\lambda_{\Sigma}\defn\sum_{m=1}^{M}\lambda_m$.  

We first consider the case $U_l=U_j$, i.e., the near user case. In this case, $\|U_j\|^2$ can be equivalently written as $\|U_j\|^2\stackrel{(d)}{=}\min\{D_1,D_2\}$ where $\stackrel{(d)}{=}$ stands for equivalence in distribution and $D_i\sim\mathrm{Exp}(\pi\lambda_{\Sigma})$ for $i=1,2$ since $U_j$ is the near user by comparing two user distances that are i.i.d. and their square has the same distribution $\mathrm{Exp}(\pi\lambda_{\Sigma})$. Hence, the pdf of $\|U_j\|^2$ can be found as
\begin{align*}
f_{\|U_j\|^2}(x) = 2\pi\lambda_{\Sigma}e^{-2\pi\lambda_{\Sigma}x},
\end{align*}
and this follows that
\begin{align}
\mathbb{E}[\exp(-s\|U_j\|^2)]&=\int_{0}^{\infty} f_{\|U_j\|^2}(x)e^{-sx}\dif x\nonumber\\
&=\frac{2\pi\lambda_{\Sigma}}{s+2\pi\lambda_{\Sigma}}\label{Eqn:Tier-mAnyUserCovProb}
\end{align}
for any $s>0$.  For the case of $U_l=U_k$, i.e., the far user case, the distribution of $\|U_k\|^2$ can be equivalently written as
\begin{align*}
\|U_k\|^2\stackrel{(d)}{=}\|U_j\|^2+D_3
\end{align*}
where $D_3\sim\mathrm{Exp}(\pi\lambda_{\Sigma})$ because $U_j$ is the nearest user and all $D_i$'s are i.i.d. exponential RVs with the memoryless property. Thus, we have
\begin{align}
\mathbb{P}[\gamma_{m,k}\geq\theta]&=\mathbb{E}\left[e^{-s(\|U_j\|^2+D_j)}\right]\bigg|_{s=\pi\lambda_{\Sigma}q\ell_m(\frac{\theta P_m}{P_{m,k}})}\nonumber\\
&=\mathbb{E}\left[e^{-s\|U_j\|^2}\right]\cdot\mathbb{E}\left[e^{-sD_3}\right]\nonumber\\
&=\left(\frac{2\pi\lambda_{\Sigma}}{2\pi\lambda_{\Sigma}+s}\right)\left(\frac{\pi\lambda_{\Sigma}}{\pi\lambda_{\Sigma}+s}\right).
 \label{Eqn:Tier-mAnyUserCovProb2}
\end{align}

Due to $\tilde{\theta}_m=\frac{\beta_m\theta}{\beta_m(1+\theta)-\theta}$ and $\beta_m\in(\frac{\theta}{1+\theta},1]$, the coverage probability of user $U_k$ can be rewritten as
\begin{align*}
\overrightarrow{\rho}_m=\mathbb{P}\left[\gamma_{m,k}\geq\tilde{\theta}_m \right].
\end{align*}
Substituting $s=\pi\lambda_{\Sigma}q\ell_m(\frac{\tilde{\theta}_mP_m}{P_{m,k}})$ and $P_{m,k}=\beta_mP_m$ into  \eqref{Eqn:Tier-mAnyUserCovProb2}  leads to the result in \eqref{Eqn:CovProbUserk}.
Next, the coverage probability in \eqref{Eqn:CovProbUj} is rewritten as
\begin{align*}
\overleftarrow{\rho}_m&=\mathbb{P}\left[\gamma_{m,j}\geq\frac{\theta(1-\beta_m)}{\beta_m(1+\theta)-\theta},\gamma_{m,j}\geq\theta\right]\\
&=\mathbb{P}\left[\gamma_{m,j}\geq(1-\beta_m)\hat{\theta}_m\right],
\end{align*}
Thus, we can obtain $\overleftarrow{\rho}_{m,l}(\beta_m,\theta)$ in \eqref{Eqn:CovProbUserj} by substituting $s=\pi\lambda_{\Sigma}q\ell_m(\theta P_m/P_{m,j})$ into   \eqref{Eqn:Tier-mAnyUserCovProb} and then replacing $P_{m,j}$ and $\theta$ in \eqref{Eqn:Tier-mAnyUserCovProb} with $(1-\beta_m)P_m$ and $\hat{\theta}_m$, respectively. 
 
\subsection{Proof of Theorem \ref{Thm:CovProbCooPNOMA}}\label{App:ProfCovProbCooPNOMA}
According to the proof of Theorem \ref{Thm:CovProbNOMA}, $\overrightarrow{\rho}_m(\beta_m,\theta)$ defined in \eqref{Eqn:CovProbCoopNOMAUk} and assuming all void BSs to be an independent PPP of intensity $(1-q)\lambda_{\Sigma}$, we know
\begin{align*}
\overrightarrow{\rho}_m(\beta_m,\theta) =\mathbb{E}\left[\exp\left(-\frac{\|U_j\|^{\alpha}}{P_m}\tilde{\theta}_m\left(I_{m,k}-\frac{S^c_m}{\theta}\right)\right)\right],
\end{align*}
\begin{align*}
\stackrel{(a)}{\approx}\
 \mathbb{E}\bigg[e^{-\pi\lambda_{\Sigma}\|U_j\|^2\left[q\ell_m\left(\tilde{\theta}_m\right)-(1-q)\ell_m\left(\frac{\tilde{\theta}_m}{\theta}\right)\right]^+}\bigg],
\end{align*}
where $(a)$ follows from the assumption that all void BSs also form an independent PPP of intensity $(1-q)\lambda_{\Sigma}$. Then by following the same approach in the proof of Theorem \ref{Thm:CovProbNOMA} to calculate the mean in the last inequality of $(a)$ for random variable $\|U_j\|^2$, the result in \eqref{Eqn:DerCoverProbCoopNOMAUserk} can be acquired.

Next, we find the accurate result of the coverage probability in \eqref{Eqn:CovProbCoopNOMAUj}. For $\beta_m\in(\frac{1+\theta}{2+\theta},1]$, we have
\begin{align*}
\overleftarrow{\rho}_m(\beta_m,\theta)\approx&\mathbb{E}\bigg[\exp\bigg(-\frac{\|U_j\|^{\alpha}}{P_m}\left(\frac{I_{m,j}}{\beta_m(1+\theta)-\theta}\right)\times\\
&\hspace{-0.5in}\max\bigg\{\theta-\frac{S^c_m}{I_{m,j}},\left(\frac{\beta_m(1+\theta)-\theta}{1-\beta_m}\right)\theta\bigg\}\bigg)\bigg| \|U_j\|\leq \|U_k\|\bigg]\\
\stackrel{(b)}{=}&\mathbb{E}\left[\exp\left(-\frac{\|U_j\|^{\alpha}I_{m,j}}{(1-\beta_m)P_m}\right)\bigg|\|U_j\|\leq \|U_k\|\right]\\
=&\mathbb{P}\left[\gamma_{m,j}\geq\theta\bigg|\|U_j\|\leq\|U_k\|\bigg|\right]
\end{align*}
where $(b)$ follows from $\frac{\beta_m(1+\theta)-\theta}{1-\beta_m}\geq 1$ due to $\beta_m\in(\frac{1+\theta}{2+\theta},1]$. Then we can use the result in \eqref{Eqn:Tier-mAnyUserCovProb} to find the approximated accurate result of  $\overleftarrow{\rho}_m(\beta_m,\theta)$ in \eqref{Eqn:DerCoverProbCoopNOMAUserj}. This completes the proof. 

% section of references

\bibliographystyle{ieeetran}
\bibliography{IEEEabrv,Ref_CoopNOMA}

\end{document}